\documentstyle[11pt,psfig]{article}
\def\reference{\parskip 0pt\par\noindent\hangindent 0.5 truecm}
\def\s{{\rm\thinspace s}}
\def\km{{\rm\thinspace km}}

\def\kms{\hbox{$\km\s^{-1}\,$}}
\def\bj{\hbox{$b_j$}} 
 
\def\mo{\hbox{M$_\odot$}}
\def\pc{{\rm\thinspace pc}}

\begin{document}
%
%
\title{Compact Stellar Systems in the Fornax Cluster:
Super-massive Star Clusters or Extremely Compact Dwarf Galaxies?
}
%


\author{M. J. Drinkwater$^{1}$ \and
  J. B. Jones$^{2}$ \and
  M. D. Gregg$^{3}$ \and
  S. Phillipps$^{4}$
} 

\date{to appear in {\em Publications of the Astronomical Society of Australia}}
\maketitle

{\center
$^1$ School of Physics, University of Melbourne, Victoria 3010,
Australia\\m.drinkwater@physics.unimelb.edu.au\\[3mm]
$^2$ Department of Physics, University of Bristol, Tyndall Avenue, Bristol, BS8 1TL, England, U.K.\\B.Jones@bristol.ac.uk\\[3mm]
$^3$        University of California, Davis, 
       and Institute for Geophysics and Planetary Physics, 
       Lawrence Livermore National Laboratory,
       L-413, Livermore, CA 94550, USA\\gregg@igpp.llnl.gov\\[3mm]
$^4$ Department of Physics, University of Bristol, Tyndall Avenue, Bristol, BS8 1TL, England, U.K.\\S.Phillipps@bristol.ac.uk\\[3mm]
}

%
\begin{abstract}

We describe a population of compact objects in the centre of the
Fornax Cluster which were discovered as part of our 2dF Fornax
Spectroscopic Survey. These objects have spectra typical of old
stellar systems, but are unresolved on photographic sky survey plates.
They have absolute magnitudes $-13<M_B<-11$, so they are 10 times more
luminous than any Galactic globular clusters, but fainter than any
known compact dwarf galaxies.  These objects are all within 30
arcminutes of the central galaxy of the cluster, NGC 1399, but are
distributed over larger radii than the globular cluster system of that
galaxy. 

We suggest that these objects are either super-massive star clusters
(intra-cluster globular clusters or tidally stripped nuclei of dwarf
galaxies) or a new type of low-luminosity compact elliptical dwarf
(``M32-type'') galaxy.  The best way to test these hypotheses will be
to obtain high resolution imaging and high-dispersion spectroscopy to
determine their structures and mass-to-light ratios. This will allow
us to compare them to known compact objects and establish if they
represent a new class of hitherto unknown stellar system.
\end{abstract}


\bigskip


\section{Introduction}

In cold dark matter (CDM) galaxy formation, small dense halos of
dark matter collapse at high redshift and eventually merge to form the
large virialised galaxy clusters seen today. The CDM model is very
good at reproducing large-scale structure, but only very recently have
the best numerical simulations (Moore et al.\ 1998) had the resolution
to trace the formation of small halos within galaxy clusters, with
masses $\approx 10^9$M$_\odot$. We do not yet know what the lower mass
limit is for the formation of halos in the cluster environment:
determining the lower limit of galaxy mass in clusters will provide an
important constraint on these models. Most of the smallest cluster
galaxies are low surface brightness dwarfs for which mass estimates
are very difficult, though comparison with field low surface
brightness dwarfs would suggest that they may be dark matter dominated
(e.g.\ Carignan \& Freeman 1988).

In this paper we describe a population of small objects we have found
in the Fornax Cluster (see also Minniti et al.\ 1998 and Hilker et
al.\ 1999) which have high surface brightness. The origin and nature
of these objects is not yet clear, but if they are a product of the
galaxy formation process in clusters, their high surface brightness
will make it possible to probe the low-mass limit discussed
above. They may represent extreme examples of compact low luminosity
(``M32-type'') dwarf ellipticals. Alternatively, these objects may be
super-massive star clusters---there is a very large
population of globular clusters associated with the central galaxy of
the Fornax Cluster, NGC 1399 (Grillmair et al.\ 1994). These objects
are generally similar to Galactic globular clusters with similar
colours and luminosities (Forbes et al.\ 1998). There is evidence that
they are not all bound to the NGC 1399 system.  Kissler-Patig et al.\
(1999) show that the kinematics of 74 of the globular clusters
indicate that they are associated with the cluster gravitational
potential rather than that of NGC 1399. They infer that the most
likely origin of these globular clusters is that they have been
tidally stripped from neighbouring galaxies. This has also been
suggested by West et al.\ (1995), although the effect would be diluted
by the large number of halo stars that would presumably be stripped at
the same time.

Bassino, Muzzio \& Rabolli (1994) suggest that the NGC 1399 globular
clusters are remnants of the nuclei of dwarf nucleated galaxies that
have survived the disrupture of being captured by the central cluster
galaxy. A related suggestion is a second model proposed by West et
al.\ (1995) that intra-cluster globular clusters could have formed in
situ in the cluster environment.  Bassino et al.\ (1994) conclude
their discussion by noting that remnant nuclei an order of magnitude
larger (and more luminous) than standard globular clusters would also
be formed in significant numbers, but that existing globular cluster
searches would not have included them.  In Section~\ref{sec-obs} of
this paper we describe how the observations of our {\em Fornax
Spectroscopic Survey} have sampled this part of the cluster population
by measuring optical spectra of all objects brighter than $B_J=19.7$
in the centre of the Fornax Cluster. In Section~\ref{sec-prop} we
describe the properties of a new population of compact objects found
in the cluster that appear to be intermediate in size between globular
clusters and the smallest compact dwarf galaxies. We discuss the
nature of these objects in Section~\ref{sec-discuss} and show that
higher resolution observations will enable us to determine if they are
more like globular clusters or dwarf galaxies.

\section{Discovery Observations: The Fornax Spectroscopic Survey}
\label{sec-obs}

Our {\em Fornax Spectroscopic Survey}, carried out with the 2dF
multi-object spectrograph on the Anglo-Australian Telescope (see
Drinkwater et al.\ 2000), is now 87\% complete in its first field to
a limit of $B_J=19.7$. The 2dF field is a circle of diameter 2 degrees
(i.e.\ $\pi$ square degrees of sky).  We have measured optical spectra
of some 4000 objects (some going fainter than our nominal limit) in a
2dF field centred on the central galaxy of the Fornax Cluster (NGC
1399). This survey is unique in that the targets (selected from
digitised UK Schmidt Telescope photographic sky survey plates) include
{\em all} objects, both unresolved (``stars'') and resolved
(``galaxies'') in this large area of sky.  The resolved objects
measured are mostly background galaxies as expected with a minor
contribution from Fornax Cluster members. The unresolved objects are
mostly Galactic stars and distant AGN, also as expected, but some are
compact starburst (and post-starburst) galaxies beyond the Fornax
Cluster (Drinkwater et al 1999a).

Finally, in addition to the dwarf galaxies already listed in the
Fornax Cluster Catalog (FCC: Ferguson 1989) which we have confirmed as
cluster members, we have found a sample of five very compact objects
at the cluster redshift which are unresolved on photographic sky
survey plates and not included in the FCC. These new members of the
Fornax Cluster are listed in Table~\ref{tab-list} along with their
photometry measured from the UKST plates.

\begin{table*}
\caption{The new compact objects
\label{tab-list}}
\center
\begin{tabular}{lllll}
\hline
Name & RA (J2000) Dec & $B_J$& $M_B$ & cz \\
     &                & (mag)& (mag) & (\kms) \\
\hline
Thales 1 & 03 37  3.30 -35 38  4.6  &19.85  & $-11.1$   &  1507   \\
Thales 2 & 03 38  6.33 -35 28 58.8  &18.85  & $-12.1$   &  1328   \\
Thales 3$^1$ & 03 38 54.10 -35 33 33.6  &17.68  & $-13.2$   &  1595  \\
Thales 4$^2$ & 03 39 35.95 -35 28 24.5  &18.82  & $-12.1$   &  1936  \\
Thales 5 & 03 39 52.58 -35 04 24.1  &19.66  & $-11.2$   &  1337   \\
\hline
\end{tabular}

Notes: (1) CGF 1-4 (2) CGF 5-4, both in Hilker et al.\ (1999) 

\end{table*}

Our 2dF measurements of unresolved objects are 80\% complete in the
magnitude range of these objects ($17.5<\bj<20.0$). There is therefore
about one more similar compact object still to be found in in our
central 2dF field. The number density of these objects is
therefore $6\pm3$ per 2dF field ($\pi$ square degrees).  Two of
the objects (the two brightest) were also identified as cluster
members by Hilker et al.\ (1999). Hilker et al.\ measured spectra of
about 50 galaxies brighter than $V=20$ in a square region of width
0.25 degrees at the centre of the Fornax Cluster. In the ``galaxies''
they include objects which were very compact, but still resolved. By
contrast our own survey covers a much larger area and also includes all
unresolved objects.

\section{Properties of the compact objects}
\label{sec-prop}

\subsection{Sizes}

\begin{figure*}
\hfil \psfig{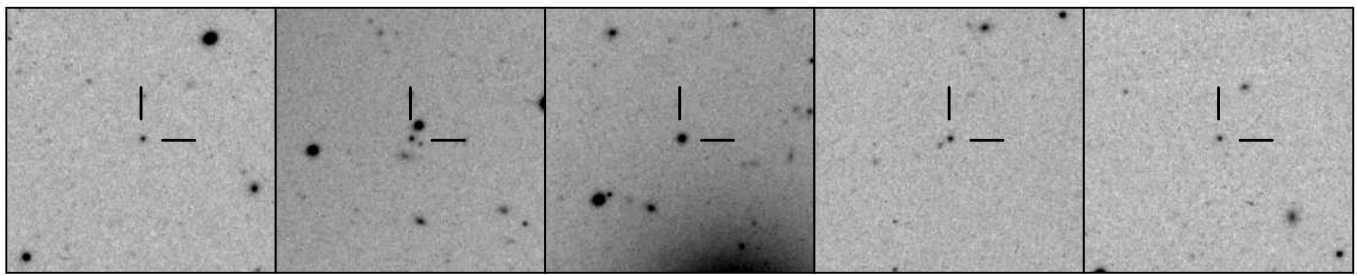}
\caption{R-band photographic images of the new compact objects. The
images are all from a single UKST exposure on Tech-Pan emulsion,
digitised by SuperCOSMOS (Miller et al.\ 1992). Each image is 2.5
arcminutes across with North at the top and East to the left.
\label{fig-image}}
\end{figure*}

These object images are unresolved and classified ``stellar'' in our
UKST plate data, although imaging with the CTIO Curtis Schmidt shows
that the brightest two objects have marginal signs of extended
structure. In Fig.~\ref{fig-image} we present R-band (Tech Pan
emulsion + OG 590 filter) photographic images of these compact objects
from the UKST. These were taken in seeing of about 2.2 arcseconds FWHM
and the third object (Thales\footnote{Thales of Miletus was the first
known Greek philosopher and scientist and possibly the earliest
astronomer.}  3) is resolved with a 3.2 arcsecond FWHM.  Applying a
very simple deconvolution of the seeing this corresponds to a scale
size (HWHM) of about 80 pc. This is much larger than any known
globular cluster, so this object, at least, is not a globular cluster.
The other objects are all unresolved, so must have scale sizes smaller
than this.

\subsection{Luminosity and Colours}

\begin{figure}
\hfil \psfig{file=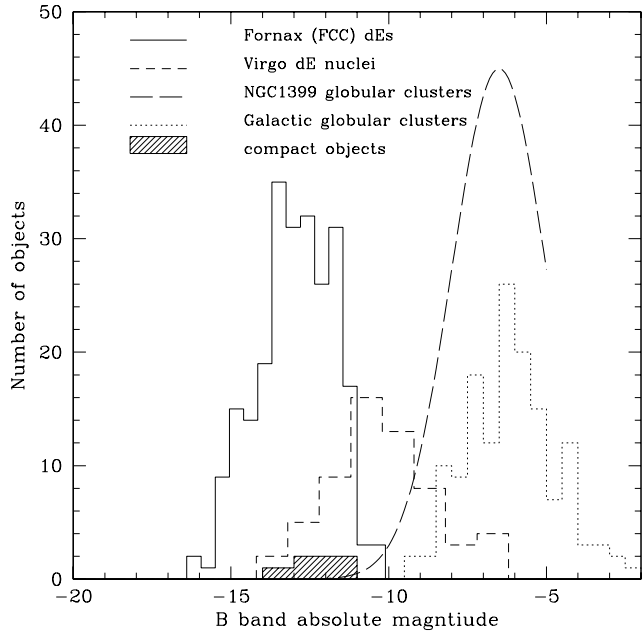,angle=0,width=9cm}
\caption{Distribution of absolute magnitude of the compact objects
(filled histogram) compared to dEs in the Fornax Cluster (Ferguson
1989; solid histogram), the nuclei of dE,Ns in the Virgo Cluster
(Binggeli \& Cameron 1991; short dashes), a model fit to the globular
clusters around NGC 1399 (Bridges, Hanes \& Harris, 1991; long dashes)
and Galactic globular clusters (Harris 1996; dotted). Note: the
magnitude limit of our survey that found the compact objects
corresponds to $M_B=-11$.
\label{fig-lf}}
\end{figure}

These new objects have absolute magnitudes $-13<M_B<-11$, based on a
distance modulus of 30.9 mag to the Fornax Cluster (Bureau et al.\
1996). These values are at the lower limit of dwarf galaxy
luminosities (Mateo 1998), but are much more luminous than any known
Galactic globular clusters (Harris, 1996) and the most luminous of the
NGC1399 globulars (Forbes et al.\ 1998) which have $M_B\approx-10$.
The luminosities of the compact objects are compared to several other
populations of dwarf galaxy and star cluster in Fig.~\ref{fig-lf}. We
note that the magnitude limit of the 2dF data corresponds to an
absolute magnitude of $M_B\approx-11$ here. In order of decreasing
luminosity the first comparison is with the dwarf ellipticals listed
in the FCC as members of the Fornax Cluster.  The possible M32-type
galaxies in the FCC are not included as none of them have yet been
shown to be cluster members (Drinkwater, Gregg \& Holman 1997). The
Figure shows that the Fornax dEs have considerable overlap in
luminosity with the compact objects, but morphologically they are very
different, being fully resolved low surface brightness
galaxies. Recently, several new compact dwarf galaxies have been
discovered in the Fornax Cluster (Drinkwater \& Gregg 1998) but these
are all brighter than $M_B=-14$ and do not match any of the objects we
discuss here.  Binggeli \& Cameron (1991) measured the luminosity
function of the nuclei of nucleated dwarf elliptical galaxies in the
Virgo Cluster. The Figure shows that this also overlaps the
distribution of the new compact objects. In this case the morphology
is the same, so the compact objects could originate from the dwarf
nuclei. The Figure also shows the luminosity functions of both the NGC
1399 globular clusters (Bridges, Hanes \& Harris, 1991) and Galactic
globular clusters (Harris 1996). These are quite similar and have no
overlap with the compact objects.

For completeness we note that the luminosities of the compact objects
have considerable overlap with the luminosities of dwarf galaxies in
the Local Group (Mateo 1998), but even the most compact of the Local
Group dwarfs, Leo I ($M_B=-11.1$) would be resolved ($r_e\approx 3''$)
in our images at the distance of Fornax. The only population they
match in both luminosity and morphology is the bright end of the
nuclei of nucleated dwarf ellipticals.

\subsection{Spectral Properties}

\begin{figure}
\hfil \psfig{file=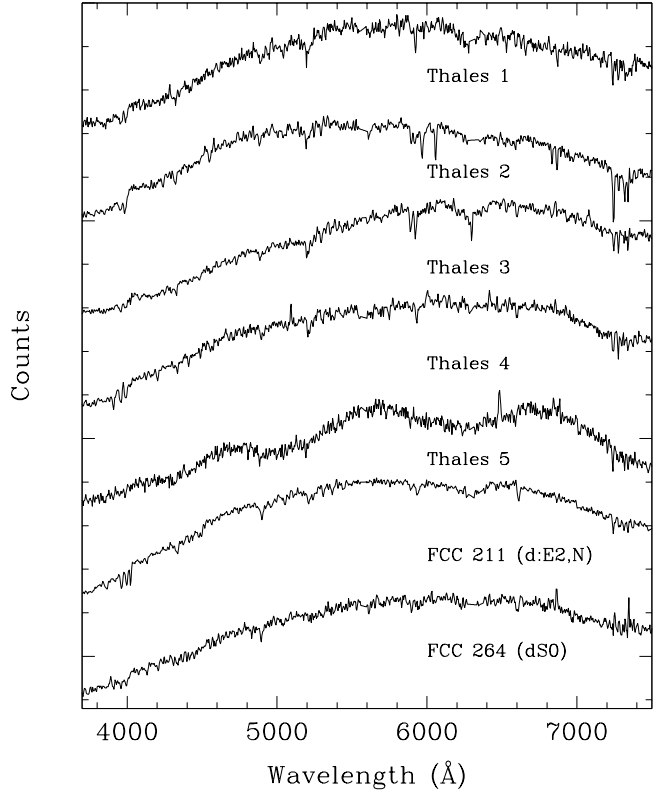,angle=0,width=9cm}
\caption{2dF discovery spectra of the five compact objects as well as
two cluster dwarf galaxies for comparison. Note: the large scale
ripple in the spectrum of Thales~5 is an instrumental effect caused by
deterioration in the optical fibre used.
\label{fig-spec}}
\end{figure}

The 2dF discovery spectra of these compact objects are shown in
Figure~\ref{fig-spec}. They have spectra similar to those of early
type dwarf galaxies in the sample (two are shown for comparison in the
Figure) with no detectable emission lines. As part of the spectral
identification process in the {\em Fornax Spectroscopic Survey}, we
cross-correlate all spectra with a sample of stellar templates from
the Jacoby et al.\ (1984) library.  The spectra of the new compact
objects were best fit by K-type stellar templates, consistent with an
old (metal-rich) stellar population. The dE galaxies observed with the
same system by contrast are best fit by younger F and early G-type
templates. This gives some indication in favour of the compact objects
being related to globular clusters, although we note that two of them
were analysed by Hilker et al.\ (1999) in more detail without any
conclusive results. We do not have the spectrum of a dE nucleus
available for direct comparison, but since our 2dF spectra are taken
through a 2 arcsec diameter fibre aperture, the spectrum of FCC 211, a
nucleated dE, is dominated by the nucleus. This spectrum was fitted by
a younger F-type stellar template, again suggestive of a younger
population than the compact objects. We cannot draw any strong
conclusions from these low-resolution, low signal-to-noise spectra.

\subsection{Radial Distribution}

\begin{figure}
\hfil \psfig{file=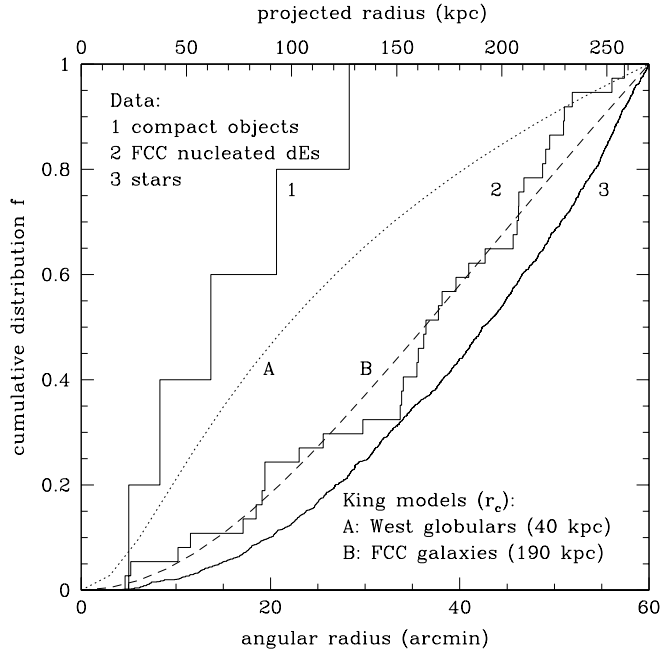,angle=0,width=9cm}
\caption{Cumulative radial distribution of the new compact objects
compared to the predicted distribution for intra-cluster globular
clusters (West et al 1995) and the profile fit to the distribution of all
FCC. Also shown is the distribution of all nucleated dwarfs in the FCC
and all the unresolved objects (stars) observed in our 2dF survey.
\label{fig-radial}}
\end{figure}

The main advantage of our survey over the previous studies of the NGC
1399 globular cluster system (e.g.\ Grillmair et al.\ 1994, Hilker et
al.\ 1999) is that we have complete spectroscopic data over a much
larger field, extending to a radius of 1 degree (projected distance of
270 kpc) from the cluster centre. This means that we can determine the
spatial distribution of the new compact objects. In
Figure~\ref{fig-radial} we plot the normalised, cumulative radial
distribution of the new compact objects compared to that of foreground
stars and cluster galaxies. This plot is the one used to calculate
Kolmogorov-Smirnov statistics and allows us to compare the
distributions of objects independent of their mean surface
densities. It is clear from the Figure that the new compact objects
are very concentrated towards the centre of the cluster, at radii
between 5 and 30 arcminutes (20--130 kpc). Their distribution is more
centrally concentrated than the King profile fitted to cluster members
by Ferguson (1989) with a core radius of 0.7 degrees (190 kpc). The
Kolmogorov-Smirnov (KS) test gives a probability of 0.01 that the
compact objects have the same distributions as the FCC galaxies: they
are clearly not formed (or acreted) the same way as average cluster
galaxies. To test the hypothesis that the compact objects are formed
from nucleated dwarfs, we also plot the distribution of all the FCC
nucleated dwarfs, as these are more clustered than other dwarfs
(Ferguson \& Binggeli 1994). However in the central region of interest
here the nucleated dwarf profile lies very close to the King profile
of all the FCC galaxies, so this does not provide any evidence for a
direct link with the new compact objects.

West et al.\ (1995) suggest that a smaller core radius should be used
for intra-cluster globular clusters (GCs). This profile, also shown in
the Figure, is more consistent with the distribution of the new
objects: the KS probability of the compact objects being drawn from
this distribution is 0.39.


\begin{figure*}
\hfil \psfig{file=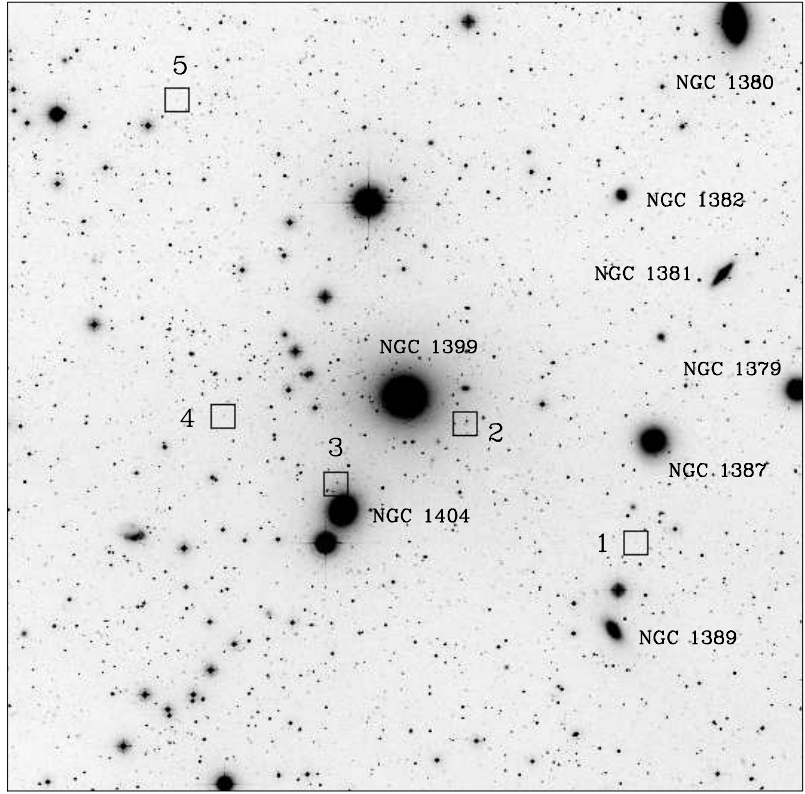,width=15cm}
\caption{The central region of the Fornax Cluster with the positions
of the new compact objects indicated by squares. This R-band
photographic image is from a single UKST exposure on Tech-Pan
emulsion, digitised by SuperCOSMOS (Miller et al.\ 1992).
\label{fig-apm}}
\end{figure*}

We also note that the radial distribution of the compact objects is
much more extended than the NGC 1399 globular cluster system as
discussed by Grillmair et al.\ and extends to three times the
projected radius of that sample. It unlikely that all the compact
objects are associated with NGC 1399. This is emphasised by a finding
chart for the central 55 arcminutes of the cluster in
Fig.~\ref{fig-apm} which indicates the location of the compact
objects. They are widely distributed over this field and Thales~3 in
particular is much closer to NGC 1404, although we note that its
velocity is not close to that of NGC 1404 (see below).

\subsection{Velocity Distribution}

We have some limited information from the radial velocities of the
compact objects. The mean velocity of all 5 ($1530\pm110 \kms$) is
consistent with that of the whole cluster ($1540\pm50 \kms$) (Jones \&
Jones (1980). However, given the small sample, it is also consistent
with the velocity of NGC 1399 ($1425\pm 4 \kms$) as might be expected
for a system of globular clusters.  Interestingly, the analysis of the
dynamics of 74 globular clusters associated with NGC 1399 by
Kissler-Patig et al.\ (1999) notes that their radial velocity
distribution has two peaks, at about 1300 and 1800\kms.
Our sample is far too small to make any conclusions about the dynamics
of these objects at present.

\section{Discussion}
\label{sec-discuss}

We cannot say much more about the nature of these objects on the basis
of our existing data. In ground-based imaging, they are intermediate
between large GCs and small compact dwarf galaxies, so it becomes
almost a matter of semantics to describe them as one or the other. The
most promising way to distinguish between these possibilities is to
measure their mass-to-light (M/L) ratios. If they are large, but
otherwise normal, GCs, they will be composed entirely of stars giving
very low M/L. If they are the stripped nuclei of dwarf galaxies we
might expect them to be associated with some kind of dark halo, but we
would not detect the dark halos at the small radii of these nuclei, so
we would also measure small M/L values. Alternatively, these objects
may represent a new, extreme class of compact dwarf elliptical
(``M32-type'') galaxy. These would presumably have formed by
gravitational collapse within dark-matter halos, so would have high
mass-to-light-ratios, like dwarf galaxies in the Local Group (Mateo
1998).  One argument against this interpretation is the apparent lack
of M32-like galaxies at brighter luminosities (Drinkwater \& Gregg
1998).  If the compact objects are dwarf galaxies, they will represent
the faintest M32-like galaxies ever found. They may also fill in the
gap between globular clusters and the fainter compact galaxies in the
surface brightness vs.\ magnitude distribution given by Ferguson \&
Binggeli (1994).

A further possibility is that these are small scale length ($\sim
100$~pc) dwarf spheroidal galaxies of only moderately low surface
brightness. While Local Group dSphs of equivalent luminosities
generally have substantially larger scale sizes (and consequently
lower surface brightnesses) (Mateo, 1998), Leo I for example has $M_B
= -11.0$, and a scale length of only 110~pc (Caldwell et al., 1992),
but as we discuss above this would be resolved in our existing
imaging.

Our existing data will only allow us to estimate a conservative upper
limit to the mass of these objects. If we say that the core radii of
the objects are less than 75\pc\ and the velocity dispersions are less
than 400\kms\ (the resolution of our 2dF spectra) we find that the
virial mass must be less than $10^{10}\mo$. For a typical luminosity
of $M_B=-12$ this implies that $M/L < 2\times 10^{3}$. This is not a
very interesting limit, so we plan to reobserve these objects at
higher spectral resolution from the ground and higher spatial
resolution with the {\em Hubble Space Telescope} (HST) in order to be
sensitive to $M/L\approx 100$. This will allow us to distinguish
globular clusters from dwarf galaxies.

\begin{figure}
\hfil \psfig{file=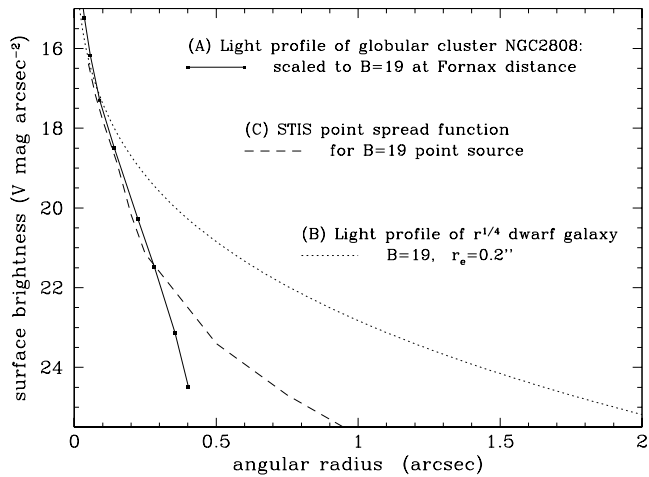,angle=0,width=9cm}
\caption{Predicted radial surface brightness profiles of the compact
objects in two extreme cases: (A) a Galactic globular cluster
(Illingworth \& Illingworth 1976) scaled 3 mag brighter in surface
brightness, and (B) a compact dwarf galaxy with an $r^{1/4}$
profile. Both are scaled to have total magnitudes of B=19 mag; they
are not corrected for instrumental PSF which is also shown
(C). \label{fig-profile}}
\end{figure}

In order to demonstrate what we could measure with high-resolution
images, we present two extreme possibilities in
Figure~\ref{fig-profile}: a very compact Galactic globular cluster and
a dwarf galaxy with an $r^{1/4}$ profile ($r_e=0.2$ arcsec), both
normalised to magnitudes of $B=19$ ($V=18.4$) and the Fornax cluster
distance.  We also plot the PSF of the {\em Space Telescope Imaging
Spectrograph} (STIS) in the Figure for reference. The globular cluster
profile is that of NGC 2808 (Illingworth \& Illingworth 1976) with the
radius scaled to the distance of the Fornax Cluster and the surface
brightness then scaled to give the desired apparent magnitude. The
globular cluster profile is very compact and will only just be
resolved with HST, but it will clearly be differentiated from the
dwarf galaxy profile.

In addition to measuring the size of these objects for the mass
measurement, the radial surface brightness profiles may also give
direct evidence for their origin and relationship to other kinds of
stellar systems. For example, if they are the stripped nuclei of
galaxies, the remnants of the outer envelope might show up in the HST
images as an inflection in the surface brightness profile at large
radius.

\section{Summary}

We have reviewed the observed properties of these new compact objects
discovered in the Fornax Cluster. Their luminosities are intermediate
between those of known globular clusters and compact dwarf galaxies,
but they are consistent with the bright end of the luminosity function
of the the nuclei of nucleated dwarf ellipticals. The 2dF spectra are
suggestive of old (metal-rich) stellar populations, more like globular
clusters than dwarf galaxies. Finally the radial distribution of the
compact objects is more centrally concentrated than cluster galaxies
in general, but extends further than the known globular cluster system
of NGC 1399.

These objects are most likely either massive star clusters (extreme
globular clusters or tidally-stripped dwarf galaxy nuclei) or very
compact, low-luminosity dwarf galaxies. In the latter case these new
compact objects would be very low-luminosity counterparts to the
peculiar compact galaxy M32. This would be particularly interesting
given the lack of M32-like galaxies at brighter luminosities
(Drinkwater \& Gregg 1998). With higher resolution images and spectra
we will be able to measure the mass-to-light ratios of these objects
and determine which of these alternatives is correct.

\section*{Acknowledgements}

We thank the referee for helpful suggestions which have improved the
presentation of this work. We wish to thank Dr.\ Harry Ferguson for
helpful discussions and for providing the STIS profile. We also
thank Dr.\ Trevor Hales for assistance in the naming of the
objects. MJD acknowledges support from an Australian Research Council
Large Grant.

\section*{References}

\reference Caldwell, N., Armandroff, T.E., Seitzer, P., Da Costa,
G.S., 1992, AJ, 103, 840
\reference Bassino, L.P., Muzzio, J.C., Rabolli, M. 1994, ApJ, 431, 634
\reference Binggeli, B.,  Cameron, L.M.,  1991, A\&A, 252, 27
\reference Bridges, T.J., Hanes, D.A., Harris, W.E., 1991, AJ, 101, 469
\reference Bureau, M., Mould, J.R., Staveley-Smith, L., 1996, ApJ,
463, 60
\reference Carignan, C., Freeman, K.C. 1988, ApJ, 332, L33
\reference Drinkwater, M.J., Gregg, M.D., 1998, MNRAS, 296, L15
\reference Drinkwater, M.J., Gregg, M.D., Holman, B.A., 1997 in
Arnaboldi M., Da Costa G.S., Saha P., eds, ASP Conf. Ser. Vol. 116,
The Second Stromlo Symposium: The Nature of Elliptical Galaxies.
Astron. Soc. Pac., San Francisco, p. 287
\reference Drinkwater, M.J., Phillipps, S., Gregg, M.D., Parker, Q.A.,
Smith, R.M., Davies, J.I., Jones, J.B., Sadler, E.M., 1999a, ApJ, 511, L97
\reference Drinkwater, M.J., Phillipps, S., Jones, J.B., Gregg, M.D.,
Deady, J.H., Davies, J.I., Parker, Q.A., Sadler, E.M., 
Smith, R.M. 2000, A\&A, submitted
\reference Ferguson H.C., 1989, AJ, 98, 367
\reference Ferguson H.C., Binggeli, B., 1994, A\&ARv, 6, 67
\reference Forbes, D.A., Grillmair, C.J., Williger, G.M., Elson,
R.A.W., Brodie, J.P. 1998, MNRAS, 293, 325
\reference Grillmair, C.J., Freeman, K.C., Bicknell, G.V., Carter, D.,
Couch, W.J., Sommer-Larsen, J., Taylor, K. 1994, ApJ, 422, L9
\reference Harris, W.E., 1996, AJ, 112, 1487
\reference Hilker, M., Infante, L., Vieira, G., Kissler-Patig, M.,
Richtler, T., 1999, A\&AS, 134, 75
\reference Illingworth, G., Illingworth, W. 1976, ApJSup, 30, 227
\reference Jacoby, G.H., Hunter, D.A., Christian, C.A., 1984, ApJSup,
56, 257
\reference Jones, J.E., Jones, B.J.T. 1980, MNRAS, 191, 685
\reference Kissler-Patig, M., Grillmair, C.J., Meylan, G., Brodie,
J.P., Minniti, D., Goudfrooij, P., 1999, AJ, 117, 1206
\reference Mateo, M., 1998, Ann. Rev. Astron. Astrophys., 36, 435
\reference Minniti, D., Kissler-Patig, M., Goudfrooij, P., Meylan, G., 1998, AJ, 115, 121
\reference Miller, L. A., Cormack, W., Paterson, M., Beard, S., Lawrence, L., 
1992, in `Digitised Optical Sky Surveys', eds. H.T. MacGillivray, 
E.B Thomson, Kluwer Academic Publishers, p. 133
\reference Moore, B., Governato, F., Quinn, T., Stadel, J., Lake, G. 1998, ApJ, 499, L5
\reference West, M.J., Cote, P., Jones, C., Forman, W., Marzke, R.O. 1995 ApJ 453 L77

\end{document}